\documentclass[12pt]{article}
\usepackage{amsfonts,amssymb}

\renewcommand{\baselinestretch}{1.4}
\newcommand{\CP}[1]{\mathbb{CP}^{#1}}

\newcommand{\R}[1]{\mathbb{R}^{#1}}

\def\Z{\mathbb{Z}}

\def\Y{\mathbf{Y}}
\def\S{\mathbf{S}}
\def\T11{{T}^{1,1}}
\def\be{\begin{equation}}
\def\ee{\end{equation}}
\def\bear{\begin{eqnarray}}
\def\eear{\end{eqnarray}}

\def\bX{\mathbf{X}}

\def\Vol{\mathrm{Vol}}
\def\vol{\mathrm{vol}}

\def\Tr{\mathrm{Tr}}

\def\bi{\bibitem}
\usepackage{graphicx}

\begin{document}

\begin{titlepage}

\begin{flushright}
hep-th/0205064\\
PUPT-2038
\end{flushright}
\vfil

\begin{center}
{\huge String Tensions and Three Dimensional}\\
\vspace{3mm}
{\huge Confining Gauge Theories}

\end{center}

\vfil
\begin{center}
{\large Christopher P. Herzog}\\
\vspace{1mm}
Joseph Henry Laboratories,  \\
Princeton University,\\
Princeton, New Jersey 08544, USA\\
{\tt cpherzog@princeton.edu}\\
\vspace{3mm}
\end{center}

\vfil

\begin{center}

{\large Abstract}
\end{center}

\noindent
In the context of gauge/gravity duality, we try to understand better 
the proposed duality between the fractional D2-brane supergravity solutions
of ({\it Nucl. Phys.} B {\bf 606} (2001) 18,
{\tt hep-th/0101096})
and a confining $2+1$ dimensional gauge theory.
Based on the similarities between this fractional D2-brane
solution and D3-brane supergravity solutions with more
firmly established gauge theory duals, we conjecture
that a confining $q$-string in the $2+1$ dimensional
gauge theory is dual to a wrapped D4-brane.  In
particular, the D4-brane looks like a string in
the gauge theory directions but wraps a 
$\S^3 \subset \S^4$ in the transverse geometry.
For one of the supergravity solutions, 
we find a near quadratic scaling law for the 
tension: $T \sim q\, (N-q)$.  Based on the 
tension, we conjecture that the gauge theory
dual is $SU(N)$ far in the infrared.  
We also conjecture
that a quadratic or near quadratic scaling
is a generic feature of confining $2+1$
dimensional $SU(N)$ gauge theories.
\vfil
\begin{flushleft}
May 2002
\end{flushleft}
\vfil
\end{titlepage}
\newpage
\renewcommand{\baselinestretch}{1.1}  


\section{Introduction}

The original AdS/CFT correspondence \cite{jthroat, US, EW}
has taught string theorists much about the relationships
between $3+1$ dimensional gauge theories and string theory
in curved space-time.  The present work is motivated
by the hope that this conjectured correspondence
generalizes to gauge theories in other dimensions.

A duality between a certain
${\mathcal N}=1$ supersymmetry preserving,
type IIA supergravity solution and a $2+1$ dimensional
confining gauge theory was proposed by 
Cveti\v{c}, Gibbons, L\"u, and Pope (CGLP) in \cite{CGLP}.
The duality was motivated by the striking resemblance 
between this supergravity solution and the Klebanov-Strassler
(KS) solution \cite{KS}, which in the infrared is the well
established dual of a confining $SU(N)$ gauge theory
is $3+1$ dimensions.  Despite the striking resemblance
of the two supergravity solutions, the $2+1$ dimensional
gauge theory dual of the CGLP solution has remained 
mysterious.  

To give the reader some context, recall that the 
KS solution grew out of an attempt to produce
a more realistic AdS/CFT correspondence.  The
original AdS/CFT correspondence,
motivated by placing a stack of $N$ 
D3-branes in flat, ten dimensional space, relates
${\mathcal N}=4$ supersymmetric $SU(N)$ gauge
theory to type IIB string theory in 
an $AdS_5 \times S^5$ background.  
To reduce the amount of supersymmetry 
to ${\mathcal N}=1$, the authors of \cite{KW} placed 
the D3-branes at the tip of the conifold, a
non-compact Calabi-Yau three-fold, 
producing an $SU(N) \times SU(N)$.  
Later,
to break the conformal symmetry, the authors
of \cite{KN,KT}, introduced $M$ fractional D3-branes,
i.e.~D5-branes wrapped on vanishingly small 
two-cycles at the tip of the conifold,
changing the gauge group to $SU(N+M)\times SU(N)$.
The original supergravity solution \cite{KT}
describing these fractional branes,
also called the Klebanov-Tseytlin or KT solution, 
had a 
singularity.  The KS solution eliminates
the singularity by deforming the conifold.
On the gauge theory side, the deformation
was related to chiral symmetry breaking
and confinement.

In an attempt to generalize some of this
work to gauge theories in other dimensions,
\cite{CGLPb, CGLP, HKfrac} produced analogs of the
KT and KS supergravity solutions for
fractional D$p$-branes, $p\neq 3$, positioned
at conical singularities.  Although most of
these supergravity solutions are well behaved
in the UV and some are well-behaved everywhere,
the gauge theory duals often remain mysterious.
In this paper,
we focus on just one of these supergravity
solutions, the fractional D2-brane
solution presented in \cite{CGLP}.
Some recent progress toward understanding
the ultraviolet region of the gauge theory was made
by \cite{LO}.  The present work is 
in large part motivated by the need to understand better the
infrared, confining region of this gauge theory.

In particular, we will use confining strings to probe
the infrared of this $2+1$ dimensional gauge theory.
In an $SU(N)$ theory, 
confining strings can be thought of as flux tubes
joining $q$ probe quarks with $q$ probe anti-quarks,
where $q$ ranges from 1 to $N-1$.  For $q=N$,
the quarks can combine into a baryon, and the
string becomes tensionless.  Moreover, there is
a symmetry under $q\rightarrow N-q$ which
corresponds to replacing quarks with anti-quarks.
Finally, one expects these confining strings to
be stable with respect to decay into strings
with fewer numbers of quarks:
\be
T_{q+q'} < T_q + T_{q'} \ .
\label{convex}
\ee
In other words, the tension should be a
convex function with respect to $q$.

For the other simple Lie groups, there are 
typically far fewer types of confining strings
because because the gluons are more effective
at screening charge.  The allowed charges
are described by the center of the Lie
group.  For $SU(N)$, the center is
$\Z_N$ while for the other 
simple Lie groups, the center is much
smaller, for 
example, $\Z_2$, $\Z_2\times \Z_2$, or $\Z_4$.
It is important to keep in mind that we don't
know for sure what the gauge group of this
$2+1$ dimensional theory is.

We propose that the confining strings are dual
to certain wrapped D4-branes in the CGLP 
supergravity solution.  The proposal is
motivated by the fact that these D4-branes
are the analog of certain wrapped
D3-branes considered by \cite{KH}
for the KS solution.
The authors of \cite{KH} argued that 
these D3-branes were the confining strings
of the ${\mathcal N}=1$ $SU(N)$ confining
gauge theory dual.

The tension of the confining strings or
equivalently of 
these wrapped D4-branes allows us to make
a conjecture about the confining 
$2+1$ dimensional gauge theory.
Treating these wrapped D4-branes as a small
perturbation on the background geometry,
we calculate the tension of the confining strings
using the DBI action.

In order to understand the conjecture, one 
should understand that there are actually two,
not just one, CGLP supergravity solutions
under discussion, and that there are as a result 
possibly two distinct $2+1$ dimensional confining
gauge theories.
Both CGLP solutions are warped compactifications
of three dimensional Minkowski space and a noncompact,
asymptotically conical, $G_2$ manifold.  The warp
factor depends on the radius associated with the
asymptotically conical limit.  One $G_2$
manifold, $\Y_S$, is a $\R{3}$ fibration over $\S^4$ while
the other, $\Y_P$, is a $\R{3}$ fibration 
over $\CP{2}$ \cite{CGLPold}.
Both solutions are stabilized by $N$ units of $F_4$ flux
that thread four cycles inside the $G_2$ manifolds.
This $F_4$ flux should correspond to $N$ 
``fractional'' D2-branes, i.e.~D4-branes wrapped
on vanishing two cyles inside the $G_2$ manifold.

The conjecture is that the confining gauge theory 
dual to the supergravity
solution involving $\Y_S$ has gauge group $SU(N)$ 
far in the infrared.
The manifold $\Y_S$ leads to a formula for the string
tensions that is nearly quadratic:
\be
T \sim q \, (N-q) \ .
\ee
By nearly, we mean that the non-quadratic corrections 
are small compared to the $q(N-q)$ term.  
The facts that $T$ is symmetric under $q \rightarrow N-q$,
that $T$ vanishes at $q=N$, and that $T$ is convex 
strongly suggest that
the gauge group is $SU(N)$.  

The quadratic or Casimir scaling of the string tensions
is interesting because it may hold true for a much larger
class of three dimensional 
confining theories than the one considered here.  
Such a quadratic or Casimir scaling hypothesis was
put forward many years ago by Ambjorn, Olesen,
and Peterson \cite{AOP}.
Lucini and Teper \cite{teper} have
also found evidence for such a quadratic, or near
quadratic, scaling for $SU(4)$ and $SU(6)$
gauge theories in $2+1$ dimensions on the lattice.\footnote{In
a calculation very similar to ours, the 
authors of \cite{CGST} 
find a quadratic scaling law for the flux tubes of a thermally
confined $SU(N)$ gauge theory in $4+1$ dimensions.}

Another line of evidence for the universality of
such a formula comes from experience with
confining strings in $3+1$ dimensional $SU(N)$
gauge theories.
Much evidence has accumulated that
these string tensions  
obey a sine law.
\be
T \sim \sin\left( \frac{\pi q}{N} \right) \ .
\label{sinelaw}
\ee
Douglas
and Shenker \cite{DS} using softly broken 
${\mathcal N}=2$ gauge theory and Hanany,
Strassler, and Zaffaroni \cite{HSZ} in the 
MQCD approach to ${\mathcal N}=1$ supersymmetric
gauge theory find this sine law.  
The wrapped D3-branes in the KS geometry \cite{KH}
obey nearly such a sine law.  In the context 
of nonsupersymmetric gauge theory, 
the authors of \cite{teper, delDebbio} using a Monte
Carlo simulation on a lattic have found 
numerical evidence for such a scaling for
non-supersymmetric
$SU(4)$, $SU(5)$, and $SU(6)$ gauge theories.  
The list goes on (see for example
\cite{Edelstein} or \cite{StrasslerQCD}).

Our second conjecture is then that this quadratic 
or near quadratic scaling is a common feature
of $2+1$ dimensional confining 
$SU(N)$ gauge theories.\footnote{
%
%
One counterexample to this near quadratic
scaling is the supergravity solution of 
Maldacena and Nastase \cite{MaNa}.  The conjectured
dual of this supergravity solution is 
2+1 dimensional ${\mathcal N}=1$ $SU(N)$ gauge theory.
However, in the far infrared, the supergravity
solution is more closely related to the KS solution
than to the $\Y_S$ fractional D2-branes.
The Maldacena-Nastase and KS solutions both become
Minkowski space cross a $\S^3$ while the $\Y_S$
solution becomes Minkowski space cross an $\S^4$.
It seems very likely that the Maldacena-Nastase
solution produces a sine law.}

The situation is less clear for $\Y_P$.  
The maximum number of confining strings
appears to be non-integer and less than
$N$. 
Additionally, the tension
formula does not vanish at $q_{max}$ and does not
have a $\Z_2$ symmetry.  
Perhaps some of the confining strings we have found
are unstable.

We begin by reviewing the CGLP supergravity
solutions in greater detail.  In the appendix,
there is an independent derivation of these
solutions from a first order system
of differential equations.  The first
order system is derived from a one dimensional effective
action.  In section 3, we present the details
of the confining string calculations.

\section{The CGLP Supergravity Solutions}

In order to extract meaningful results from
the confining string calculation in 
section 3, we need to normalize carefully
the two CGLP supergravity solutions
\cite{CGLP}.  
In the same way that the normalizations
found in \cite{HKO} laid the groundwork
for the confining string calculations
in \cite{KH}, this section lays
the groundwork for the tensions
we calculate in the next.
So the reader 
can understand where the normalizations
come from, in the following section
we have
reproduced some of what can be found 
in \cite{CGLP} or \cite{CGLPold}.

The two CGLP supergravity solutions  
under investigation
are closely related to the type IIA supergravity
solution corresponding to a stack of D2-branes
placed in flat ten dimensional space.
The CGLP solutions
are built out of a warped compactification of $2+1$ dimensional
Minkowski space and an asymptotically conical $G_2$ manifold
$\Y$.  Hence the metric is
\be
ds_{10}^2 = H(r)^{-5/8} dx^\mu dx^\nu \eta_{\mu\nu} + 
H(r)^{3/8} ds_\Y^2 \ .
\ee
The variable $r$ is the radius of the asymptotically conical
region of $\Y$.  

The metric of $\Y$ can be written down for arbitrary $r$.  In both
cases, $\Y$ is an $\R{3}$ bundle over a four dimensional Einstein manifold
$M_4$.  
\be
ds_\Y^2 = \ell^2 \left[ h(r)^2 dr^2 + a(r)^2 (D\mu^i)^2 + b(r)^2 ds_{M_4}^2 
\right]\ .
\label{g2metric}
\ee
The coordinates on $\Y$ are all dimensionless, so the length scale
$\ell$ has been introduced to give the metric the right scaling.
In one case, $\Y=\Y_S$ and  $M_4= \S^4$, and in the other, $\Y=\Y_P$ and
$M_4 = \CP{2}$.  The Einstein condition on $M_4$ is such that
$R_{ab} = 3 g_{ab}$.  Now, the $\mu^i$ are coordinates on the
$\R{3}$ subject to $\mu^i\mu^i=1$.  The fibration is written
in terms of $SU(2)$ Yang-Mills instanton potentials $A^i$ where
\be
D\mu^i = d\mu^i+\epsilon_{ijk}A^j \mu^k \ .
\ee
The field strengths $J^i \equiv dA^i + \frac{1}{2} \epsilon_{ijk}A^j \wedge A^k$
satisfy the algebra of the unit quaternions
$J_{\alpha\gamma}^i J_{\gamma\beta}^j = -\delta_{ij} \delta_{\alpha \beta}
+ \epsilon_{ijk} J_{\alpha \beta}^k$.

The metric (\ref{g2metric}) 
is Ricci-flat and has $G_2$ holonomy when the functions
$h$, $a$, and $b$ satisfy
\be
h^2 = \left(1-\frac{1}{r^4} \right)^{-1} \; , \; \; \;
a^2 = \frac{1}{4} r^2 \left(1 - \frac{1}{r^4} \right) \; , \; \; \;
b^2 = \frac{1}{2}r^2 \ .
\label{hab}
\ee
The variable $r$ runs from one to infinity.  Performing
the rescaling $R= r \ell$, one sees that this parameter
$\ell$ is very similar to the deformation parameter
$\epsilon$ of the deformed conifold \cite{KS}.
At the tip $r=1$ of this deformed cone, the $\R{3}$ directions
vanish while the $M_4$ remains finite:
\be
ds_\Y^2 \rightarrow \frac{\ell^2}{2} ds^2_{M_4} \ .
\ee
In the other limit $r\rightarrow \infty$, the metric approaches that 
of a cone over a squashed Einstein manifold.
The metric 
separates into two pieces:
\be
ds_\Y^2 = \ell^2 \left[ dr^2 + r^2 ds_\bX^2 \right]
\ee
where $\bX$ is a six dimensional ``squashed'' 
Einstein manifold satisfying $R_{ab} = 5 g_{ab}$.
In particular, this squashed metric is nearly K\"ahler, a technical 
condition with some unsettling consequences (see, for example \cite{Acharya}).
For example, the first Chern class vanishes on nearly K\"ahler spaces even
though the Ricci curvature does not always. 
For the case $\Y_S$, the Einstein manifold is a squashed
$\CP{3}$ while for $\Y_P$, the Einstein manifold is the nearly
K\"ahler flag manifold $SU(3) / T^2$, where $T^2$ is a maximal
torus in $SU(3)$.

There are also fluxes that stabilize the metric.  The four form
RR flux $F_4$ has two pieces, one corresponding to the electric
flux of the ordinary D2-branes aligned in the Minkowski
space-time directions, the other corresponding to magnetic flux
from the ``fractional'' D2-branes.  These fractional D2-branes
are D4-branes wrapped on vanishing 2-cycles inside $\Y$.  As a 
result, they source a flux through a transverse
4-cycle inside $\Y$:
\be
g_s F_4 = d^3x \wedge dH^{-1} + m G_4 \ .
\ee
Just as in the standard D2-brane solution where $m=0$, the
dilaton is nonzero, $e^\phi = g_s H^{1/4}$.

Finally, to satisfy the supergravity equations of motion,
a nonzero $m$ forces one to turn on the NSNS three form
flux
\be
H_3 = m G_3 \ ,
\ee
where $G_3$ is a harmonic 3-form inside $\Y$ and
$G_4 = \star G_3$ with $\star$ the Hodge dual with respect
to $ds_\Y^2$.
The trace of Einstein's equations enforces the condition
on the warp factor
\be
\label{LapH}
\square H = -\frac{1}{6} m^2 |G_3|^2 \ 
\ee
where $\square$ is the Laplacian with respect to $ds_\Y^2$
and the magnitude $|\cdots|^2$ is also taken with 
respect to $ds_\Y^2$.

The harmonic 3-form $G_3$ is 
\be
\ell G_3 = f_1(r) \, dr \wedge X_2 + f_2(r) \, dr \wedge J_2 + f_3(r) \, X_3
\ee
where
\be
X_2 \equiv \frac{1}{2} \epsilon_{ijk} \mu^i D\mu^j \wedge D\mu^k \; , \; \; \;
J_2 \equiv \mu^i J^i \; , \; \; \; 
X_3 \equiv D\mu^i \wedge J^i \ .
\ee
The forms obey
\be
dX_2 = X_3 \; , \; \; \; dJ_2 = X_3 \; , \; \; \; dX_3 = 0 \ .
\ee
Using these relations and the fact that $f_3' = f_1 + f_2$, one can 
choose a gauge such that
\be
\ell B_2 = m\left(\int_1^r f_1(u) du \right) X_2 + 
m\left(\int_1^r f_2(u) du \right) J_2 
\ee
and $dB_2 = H_3$.  Note that at $r=1$, $B_2$ vanishes.

The dual four form $G_4$ is
\be
G_4 = f_3 h \, \epsilon_{ijk} \mu^i dr \wedge D\mu^j \wedge J^k + 
\frac{f_2 a^2}{h} X_2 \wedge J_2 + 
\frac{f_1 b^4}{2ha^2} J_2 \wedge J_2 \ .
\ee
One then solves for the $f_i$ using the fact that $dG_4=0$ 
and that $G_4$ should be responsible for a constant
flux from the the fractional D2-branes.  The result is 
\be
f_1 = ha^2 u_1 \; , \; \; \; f_2 = hb^2 u_2 \; , \; \; \;
f_3 = ab^2 u_3 \ ,
\ee
where 
\begin{eqnarray}
u_1 &=& \frac{1}{r^4} + \frac{P(r)}{r^5(r^4-1)^{1/2}} \ , \nonumber \\
u_2 &=& -\frac{1}{2(r^4-1)} + \frac{P(r)}{r(r^4-1)^{3/2}} \ , \\
u_3 &=& \frac{1}{4r^4(r^4-1)} - \frac{(3r^4-1) P(r)}{4r^5(r^4-1)^{3/2}} \ ,
\nonumber \\
\end{eqnarray}
and where
\be
P(r) = \int_1^r \frac{d\rho}{\sqrt{\rho^4-1}} \ .
\ee

We will now attempt to quantize $G_4$ flux by looking at the limit
$r\rightarrow 1$. 
In this limit, the $u_i$ all become constant:
\be
u_1 \rightarrow \frac{3}{2} \; , \; \; \;
u_2 \rightarrow -\frac{1}{4} \; , \; \; \;
u_3 \rightarrow -\frac{1}{4} \; .
\ee
However, the choice of radial variable introduces a coordinate singularity
to the metric because 
\be
h^2 dr^2 \rightarrow \frac{dr^2}{4(r-1)} \ .
\ee
It would be best to introduce a new radial variable $\tau$ such that
$d\tau \sim dr / \sqrt{r-1}$.  However, we will just pretend that
$h \, dr$ is well behaved in the limit $r=1$.  Putting the pieces
together, in the limit $r \to 1$,
\be
\label{Gone}
G_4  \rightarrow \frac{3}{8} \vol(M_4) \ ,
\ee
where we have used the fact 
that $J_2 \wedge J_2 = 2 \vol(M_4)$.\footnote{
In our notation
\[
\Vol(\bX) = \int_\bX \vol(\bX) \ .
\]}

From the Dirac quantization condition, we know what the integral
of $F_4$ over the $M_4$ at the tip of the cone should be
\be
\int_{M_4} F_4 = 8 \pi^3 \alpha'^{3/2} N
\ee
where $N$ is the number of fractional D2-branes.
As the volume of a unit $\S^4$ is $8\pi^2/3$ and the 
volume of our $\CP{2}$ is $2\pi^2$, one finds that
for the manifold $\Y_S$, $m_S = 8 \pi \alpha'^{3/2} g_s N$
while for $\Y_P$, $m_P = \frac{32}{3} \pi \alpha'^{3/2} g_s N$.

One might wonder if this number changes as the radius changes.
To allay these fears, note that
\be
d(\epsilon_{ijk} 
\mu^i D\mu^j \wedge J^k) = 2 X_2 \wedge J_2 + 2 J_2 \wedge J_2 \ .
\ee
As a result, one can write for arbitrary $r$ that  
\be
G_4 = \frac{3}{16} J_2 \wedge J_2 + 
d \left( (g_3(r)+c_2) \epsilon_{ijk} 
\mu^i D\mu^j \wedge J^k \right) 
\ee 
where $g_3' = f_3 h$ and $c_2$ is a constant.  
Thus the limit (\ref{Gone}) is related
to the general expression for $G_4$ by an exact form. 

Next, we calculate the warp factor.
We could solve the second order 
differential equation
(\ref{LapH}).  However, because
our solution preserves ${\mathcal N}=1$
supersymmetry, it should not be altogether
surprising that the warp factor $H(r)$
can be derived from a first order 
differential equation.
Consider the equation of motion for
the four form $F_4$,
\be
d \star \left( e^{\phi/2} F_4 \right) =
-g_s^{1/2} F_4 \wedge H_3 \ .
\ee
This equation is equivalent to (\ref{LapH}).
Because of the properties of the fluxes,
it turns out we can integrate this
equation to get
\be
\star \left( e^{\phi/2} \, d^3x \wedge dH^{-1} \right) = 
\left. -g_s^{-1/2} m G_4 \wedge B_2 \right|_{M_6} \ .
\label{onedwarp}
\ee
The differential
forms are restricted to lie on $M_6$, the level
surfaces of the $G_2$ manifold.  
We have set an integration constant to
zero.  Physically, setting this constant to
zero eliminates our freedom to choose the
amount of flux from the ordinary D2-branes.
Another choice for this flux would
result in an IR singularity.

In the appendix, 
using a one dimensional effective action,
we derive a complete system
of first order differential equations 
that describes this fractional D2-brane solution.
This effective action method provides an 
alternative derivation of (\ref{onedwarp}).

After some algebra, (\ref{onedwarp})
can be solved to yield
\be
H(r) = \frac{m^2}{2 \ell^6} \int_r^\infty \rho 
\left( 2u_2(\rho) u_3(\rho) - 3u_3(\rho) \right) d\rho \ .
\label{Hformula}
\ee
The integration constant
has been chosen such
that $H(r) \sim \frac{Q}{r^5}$ in the limit
$r\rightarrow \infty$.  In other words, 
we have dropped the 
asymptotically flat part of the metric,
thus
taking the near throat limit,
zooming in on the gauge theory dynamics.

In the other limit, $r=1$,
\be
H(1) = \frac{m^2}{\ell^6} a_0
\ee
where
\be
a_0 \equiv 
\int_1^\infty \rho 
\left( 2u_2(\rho) u_3(\rho) - 3u_3(\rho) \right) d\rho \approx 
0.10693\ldots \ .
\ee
The fact that $H(r)$ becomes a constant at small
$r$ was the original reason motivating the belief
that the gauge theory dual is confining.
Another choice for the ordinary D2-brane flux
in (\ref{onedwarp}) would have caused $H(r)$ to
diverge at $r=1$.

For completeness, we also 
normalize the number of
ordinary D2-branes although
we won't need this number in the following.  
This number can be obtained
from the four form $F_4$:
\be
\star \left( e^{\phi/2} d^3x \wedge dH^{-1}\right) = 
H' r^6 \ell^5 \vol(M_6)
\ee
where $M_6$ is a level surface, $r=const$, of
$\Y$.  
The Dirac quantization condition
enforces that 
the number of D2-branes $N_2$ 
satisfy
\be
\int_{M_6} \star 
\left(e^{\phi/2} d^3x \wedge dH^{-1} \right) = 
32 \pi^5 \alpha'^{5/2} g_s N_2 \ .
\ee

Now in general, $H(r)$ varies with respect
to $r$, so the number of D2-branes will
vary as well.  However, at large
$r$, 
\be
H(r) \rightarrow \frac{Q}{r^5} - \frac{m^2}{4 \ell^6 r^6} + \ldots
\ee
The asymptotic value of $N_2$ is thus 
\be
Q = \frac{32 \pi^5 \alpha'^{5/2}}{5 \ell^5 \Vol(M_6)} g_s N_2 \ .
\ee
Moreover, from (\ref{Hformula}), it is clear that there
 is a relation between $Q$ and $m^2$.  In particular
\be
Q = \frac{2m^2}{5\ell^6} b_0 
\ee 
where 
\be
b_0 = \frac{9\sqrt{2}}{32} {\mathbf K}\left( \frac{1}{\sqrt{2}} \right) 
= 0.73745\ldots \ 
\ee
and ${\mathbf K}(k)$ is a complete elliptic integral of the first kind
\be
{\mathbf K}(k) = \int_0^{\pi/2} \frac{d\theta}{\sqrt{1-k^2 \sin^2 \theta}} \ .
\ee

\section{Confining Strings}

The hypothesis is that confining strings
in the gauge theory duals of these
type IIA supergravity backgrounds are
dual to certain wrapped D4-branes.
In particular, these D4-branes
have $q$ units of electric
flux in the $01$ directions,
indicating the presence of $q$
dissolved fundamental strings.
At the tip of the
cone $r=1$, a four manifold $M_4$ remains
of finite size.  These D4-branes
can wrap a topologically trivial three
cycle in $M_4$.  The wrapping is then
stabilized by the four form
flux $F_4$ threading $M_4$ at the tip of
the cone.  The remaining $(1+1)$ dimensions
of the D4-brane constitute a 
confining $q$-string in the flat, $2+1$ dimensional
transverse space.  The tension of these
strings is roughly proportional to the volume
of the wrapped three cycle.

The reason for such a guess comes from the 
KS solution \cite{KS} where the 
geometry is very similar but the
gauge theory dual much more well
established.  In the KS solution,
at the tip of the cone, one is
left with a finite $\S^3$ and flat
$3+1$ dimensional Minkowski space.
In this geometry, confining 
strings were dual to wrapped D3-branes.  The
D3-branes wrapped an $\S^2$ inside
the $\S^3$ and were stabilized
by a three form flux threading
the $\S^3$ \cite{KH}.

Another intuitive way of understanding
the hypothesis is through the Myers 
effect \cite{Myers}.  A priori, it
seems a natural guess to associate
fundamental strings with the 
confining strings of a gauge theory.
However, in the presence of a background 
RR field, such a stack of strings can
``blow up'' into a D-brane of higher
dimension.

To investigate our hypothesis, we 
will use a probe approximation,
assuming the extra wrapped D4-brane
produces a negligible back-reaction on
the metric.  In particular,
our D4-brane action is
\be
S = -\mu_4 \int d^5x \, e^{-\phi} 
\sqrt{-\det(G_{ab} + 2 \pi \alpha' {\mathcal F}_{ab})}
+
2\pi \alpha' \mu_4 \int C_3 \wedge {\mathcal F}_2
\label{DBI}
\ee  
where the integral is over the D4-brane world-volume,
namely the $01$ directions and a three cycle inside
$M_4$.  The 
two form ${\mathcal F}_2$ is the
gauge field on the D4-brane.  The tension $\mu_4$ is
\be
\mu_4 = \frac{1}{16 \pi^4 \alpha'^{5/2}} \ .
\ee 
We expect the probe calculation to be valid in the large
$N$ limit, where the number of background D4-branes
is large compared to the single probe brane.

The DBI action is written in terms of the 
string frame metric.
In the limit $r\rightarrow 1$, 
the full ten dimensional, string frame metric reduces to
\be
ds_{10}^2 \rightarrow
H(1)^{-1/2} dx^\mu dx^\nu \eta_{\mu\nu} +
H(1)^{1/2} \frac{\ell^2}{2} ds_{M_4}^2 \ .
\ee
The calculations are sufficiently different for the
$M_4$ that we will cover the $\S^4$ and
$\CP{2}$ cases separately.

\subsection{Confining Strings and $\S^4$}

We use the standard metric on $\S^4$:
\be
ds_{\S^4}^2 = d\psi^2 + \sin(\psi)^2 ds_{\S^3}^2 \ .
\ee
Our D4-brane will sit at a constant $\psi$ and wrap
the corresponding $\S^3$.  In these coordinates,
at $r=1$,
\be
C_3 = \frac{d^3x}{H(1)} + \frac{3m}{8g_s} \xi(\psi) \vol(\S^3) \ ,
\ee
where
\be
\xi(\psi) \equiv \int_0^\psi \sin(u)^3 du \ ,
\ee
and hence 
\be
F_4 = \frac{3m}{8g_s} \vol(\S^4) \ .
\ee
The probe D4-brane will only be sensitive to the
$\S^4$ dependent piece of $C_3$.  

To get an effective Lagrangian for the probe brane,
we integrate (\ref{DBI}) over the $\S^3$ and 1
directions.  To keep the action finite and to make
the quantization easier, it is convenient to 
make the 1 direction periodic with length $\lambda$.
The effective Lagrangian describing the wrapping is then
\be 
{\mathcal L} = -A \sqrt{1-E^2} \sin(\psi)^3 + QE\xi(\psi)
\ee
where for convenience, we have defined  
\be
E \, dx^0\wedge dx^1 \equiv 2 \pi \alpha' H(1)^{1/2} {\mathcal F} \ . 
\label{Edef}
\ee
The constants $A$ and $Q$ are 
\be
A = \frac{\mu_4 \lambda \ell^3}{g_s} \frac{\pi^2}{\sqrt{2}} \ , \; \; \;
Q = \frac{\mu_4 \lambda \ell^3}{g_s} \frac{3 \pi^2}{4a_0^{1/2}} \ .
\ee

We choose a gauge in which ${\mathcal F}_{01} = \dot{A_1}$.
The existence of large gauge transformations means
that $A_1$ is periodic with period $2\pi /\lambda$.  Thus, 
the conjugate variable
\be
\frac{\partial {\mathcal L}}{\partial \dot{A_1}} 
\ee
is quantized in units of $\lambda$, and
\be
\frac{\partial {\mathcal L}}{\partial E} \equiv P 
= \frac{q\lambda}{2\pi\alpha' H(1)^{1/2}}
\label{quantP}
\ee
where $q$ is an integer \cite{CK}.

The wrapped D4-brane is stable when the
energy is minimized.  The first step is
to write a Hamiltonian for the system:
\be
H = PE - {\mathcal L} = 
\sqrt{A^2 \sin(\psi)^6 + (P-Q\xi)^2} \ .
\ee
Next we minimize $H$ with respect to $\psi$.
Solving $dH/d\psi = 0$, there is a critical
point when $\psi$ satisfies
\be
\frac{P}{Q} = \int_0^\psi \sin(u)^3 du + \frac{A^2}{Q^2} (\sin(\psi)^3)' \ .
\label{PQ}
\ee
Indeed, this critical point is a minimum of the energy provided
$A^2/Q^2 < 1/3$.

Substituting this expression back into the Hamiltonian yields
\be
H_{min} = \lambda T = A \sin(\psi)^2 \left[ \sin(\psi)^2 + 
\left(\frac{3A}{Q}\right)^2 \cos(\psi)^2 \right]^{1/2} \ .
\label{Hmin}
\ee
This energy divided by $\lambda$ is precisely the tension $T$ 
of our confining strings.  The variable $\psi$ is
to be interpreted as a function of $P$ using
(\ref{PQ}).  For general values
of $A/Q$, (\ref{PQ}) is a cubic
polynomial in $\cos(\psi)$.  Thus
in principle we have an analytic
albeit messy expression for $\psi(P)$.
In turn, the momentum $P$ is related to
the number of units $q$ of quantized electric
flux on the wrapped D4-brane via (\ref{quantP}), and hence to 
the type of confining string.  In short,
(\ref{Hmin}) should be interpreted as a function
of $q$, the number of probe quarks at the ends
of the confining string.

The expression $H_{min}$ has the features one would expect
of the tension of a confining string in an $SU(N)$ gauge
theory.  First the tension vanishes when the number
of confining strings $q=0$ and when $q=N$.  Indeed, 
the tension clearly vanishes at $\psi=0$, where $P=0$, and
at $\psi = \pi$ where $P=4Q/3$.  From the quantization condition
on $P$ and the expression for $Q$, 
one can see that 
$P=4Q/3$ corresponds to $N$ quanta of flux 
dissolved in the D4-brane whereas $P=0$ corresponds
to no dissolved flux.

Second, the tension is symmetric under $q \to N-q$,
reflecting the freedom to replace quarks with
anti-quarks without changing the tension.  Clearly
$H_{min}$ is symmetric under $\psi \to \pi - \psi$.
Inspection of (\ref{PQ}) reveals that 
under $\psi \to \pi - \psi$, $P/Q$ becomes $4/3 - P/Q$, or
in other words $q \to N-q$.

Third, the tension is a convex function of $P$ for
the region of parameters where $A^2/Q^2 < 1/3$.
Indeed, if we are sitting at a minimum of
$H$ with respect to $\psi$, $H$ must be convex
with respect to $P$ because
\be
\frac{\partial^2 H}{\partial P^2} 
\left( \frac{\partial P}{\partial \psi_0} \right)^2 = 
\left. \frac{\partial^2 H}{\partial \psi^2} \right|_{\psi=\psi_0}
\ee 
where $\psi_0$ is the location of the minimum.
In other words (\ref{convex}) is valid and a confining
$q$-string is stable with respect to decay into
confining strings with smaller $q'<q$. 
Indeed, for us, 
\be
\frac{A}{Q} = \frac{4}{3\sqrt{2}} a_0^{1/2} = 0.30830\ldots \ .
\ee

Based on these three pieces of evidence, it seems
very likely that the confining gauge theory dual
has gauge group $SU(N)$.  One might
wonder about the existence of a Chern-Simons term
in the gauge theory action:
\be
S_{CS} = \frac{k}{4\pi} \int \Tr(A \wedge dA + \ldots) \ .
\ee
The solution preserves ${\mathcal N}=1$
supersymmetry.  Using the usual
Witten index argument, $k$ is expected
to be at least $N/2$.  
According to \cite{Wittenindex},
an area law for Wilson loops in 
the gauge theory  
means that $k \leq N/2$;
our confining strings are essentially 
Wilson loops.
We conclude that there
is a Chern-Simons term and that
$k = N/2$.\footnote{
To be more precise, classically
the Chern-Simons term has a 
coefficient of $N/2$.  Quantum 
corrections reduce the coefficient
to zero.}

Finally, note that the expression $H_{min}$ simplifies
remarkably, becoming quadratic, if $A/Q = 1/3$, for in this case
(\ref{PQ}) becomes linear in $\cos \psi$,
\be
\frac{P}{Q} = \frac{2}{3}(1-\cos\psi)
\ee
and
\be
H_{min} = \lambda T = \frac{4A}{N^2} q \left(N- q\right) \ .
\ee
In our case, $A/Q \approx 1/3$, and so the tension
is arguably approximately quadratic.
Indeed, if $H_{min}$ is plotted as a function
of $q$ for the two cases,
$A/Q = 1/3$ and $A/Q = 0.30830\ldots$,
the two curves are virtually
indistinguishable (see Figure 1).
Based on experience with
the $3+1$ dimensional cases,
we conjecture that a quadratic
scaling is a common feature of 
theories in the universality class of
confining $2+1$ dimensional $SU(N)$ 
gauge theory.

\begin{figure}
\includegraphics[width=4in]{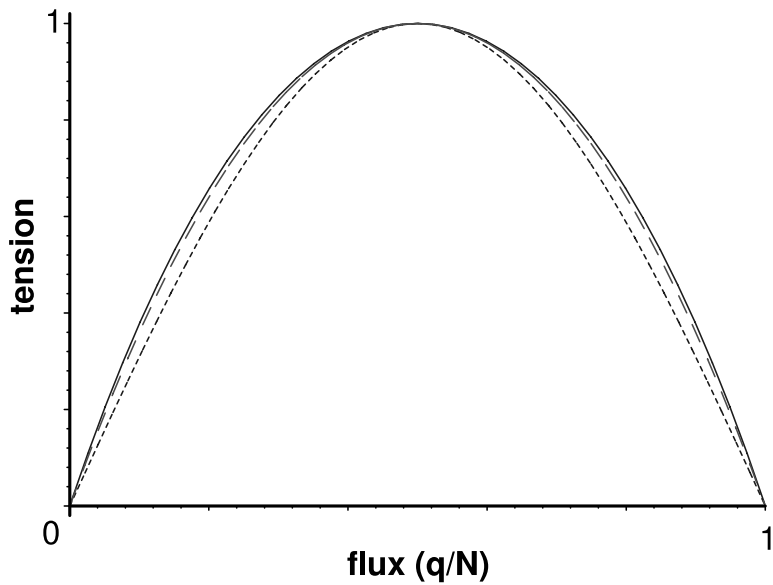}
\caption{A plot of the $q$-string tension versus $q/N$.  The solid,
outermost line is the tension computed using the 
$\Y_S$ CGLP fractional 
D2-brane model.  The dashed line corresponds to 
quadratic scaling, $q\, (N-q)$.  The dotted line,
farthest inside, is a sine law. 
The maximum tensions are normalized to one.}
\label{fig1}
\end{figure}

\begin{figure}
\includegraphics[width=4in]{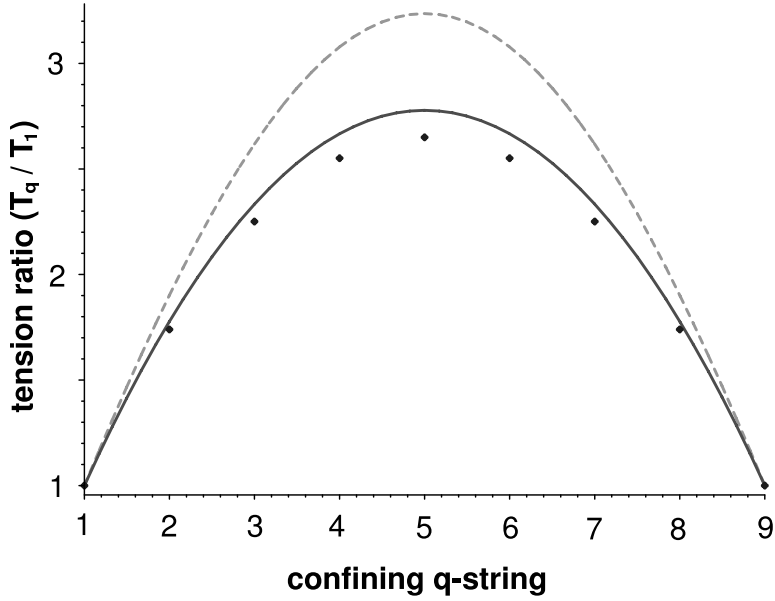}
\caption{A plot of $T_q / T_1$ versus $q$ for
an $SU(10)$ gauge theory.  The diamonds
were calculated using the $\Y_S$ fractional D2-brane model.
The solid line corresponds to a quadratic scaling law.
The dashed line comes from the sine scaling law.
}
\label{fig2}
\end{figure}

\subsection{Confining Strings and $\CP{2}$}

Although the $\S^4$ confining strings 
worked amazingly well, the confining
strings involving $\CP{2}$ remain
murky.  To try to give
a balanced account, we present our results
for this more ambiguous case.

We choose coordinates on $\CP{2}$ where the 
Einstein metric is
\be
ds^2= 2\left(d\mu^2 + \frac{1}{4} \sin(\mu)^2 
( \sigma_1^2 + \sigma_2^2 + \cos(\mu)^2 \sigma_3^2 ) \right)
\ee
and where
\be
\sigma_1 = d\theta \; ; \; \; \;
\sigma_2 = \sin(\theta) d\phi \; ; \; \; \;
\sigma_3 = (d\psi + \cos(\theta)d\phi) \ .
\ee
The coordinates $(\psi, \theta, \phi)$ are the Euler angles on a $\S^3$.
We make an ansatz where 
the D4-branes wrap this squashed $\S^3$ sitting inside
$\CP{2}$.  At the end, we will minimize the energy as a function
of $\mu$.  While rotational invariance of the $\S^4$ made choosing
a three cycle relatively easy, it is not completely clear that
this $\S^3 \subset \CP{2}$ is the best, in the sense of most
stable, submanifold to choose.

The Lagrangian is
\be
{\mathcal L} = -A \sin(\mu)^3 \cos(\mu) \sqrt{1-E^2} + QE \sin(\mu)^4 \ 
\ee
where
\be
A = \frac{\mu_4 \lambda \ell^3}{g_s}  2\pi^2 \ , \; \; \;
Q = \frac{\mu_4 \lambda \ell^3}{g_s} \frac{3\pi^2}{4a_0^{1/2}} \ .
\ee 
The electric field $E$ is defined as in
(\ref{Edef}).

There is also a conserved  
momentum $P = d{\mathcal L}/dE$ quantized 
exactly as above.  The Hamiltonian for the system
is then
\be
H = \sqrt{A^2 \sin(\mu)^6 \cos(\mu)^2 + (P-Q\sin(\mu)^4)^2} \ .
\ee

To minimize the Hamiltonian, one finds that $\mu$ must
satisfy
\be
(P-Q\sin(\mu)^4) = \frac{A^2}{Q} \left(
\frac{3}{4} \sin(\mu)^2 - \sin(\mu)^4 
\right) \ .
\ee
This value of $\mu$ is indeed a minimum
provided $A^2/Q^2 < 8/5$.
Substituting this expression back into $H$, one
finds that the minimum energy is
\be
H_{min} = 
A \sin(\mu)^2 \left[
\left(\frac{A^2}{Q^2} -1 \right) \sin(\mu)^4
+ \left( 1- \frac{3A^2}{2Q^2} \right) \sin(\mu)^2 + 
\frac{9A^2}{16Q^2}  \right]^{1/2}\ .
\ee
The range of $\mu$ is $0 \leq \mu \leq \pi/2$.  Thus
we see that while $H_{min}$ vanishes at $\mu=0$, it does 
not do so at $\mu = \pi/2$, precluding any 
$\Z_2$ symmetry.

This example is stranger still.
The maximum value of $P/Q$ is 
\be
\left(\frac{P}{Q}\right)_{max} = 1 - \frac{A^2}{4Q^2} \ .
\ee
As $A/Q = 8 a_0^{1/2} /3 = 0.87201\ldots $, the maximum number
of confining $q$-strings allowed is
\be
q_{max} = N \left(1- \frac{A^2}{4Q^2}\right) \ ,
\ee
some noninteger number\ldots.

The problem may lie with our choice
of a submanifold of $\CP{2}$.  We hope to
return to this example in the future, armed
with a better ansatz.

\section*{Acknowledgments}
I would like to thank Igor Klebanov for 
giving me the idea to write this paper.  I would also
like to thank Chris Beasley, Aaron Bergman, and Peter
Ouyang for discussions.  
I am grateful to Chris Pope for correspondence.
This work was partially
supported by NSF Grant PHY-9802484.

\begin{appendix}

\section{First Order System}

As a check of the fractional D2-brane solutions
presented in section 2, we now
present an independent derivation that
makes use of an
effective one dimensional action.  
We will see that this one dimensional
action can be written in terms of a 
superpotential.  Moreover, the
metric and field strengths of the previous
section derive from a first order
system of differential equations arising
from this superpotential.
That the fractional D2-brane solution 
obeys a first order system of equations
strengthens the claim of \cite{CGLP}
that this supergravity solution
does indeed preserve supersymmetry.

The idea is to
start with the full type IIA supergravity
Lagrangian
\begin{eqnarray}
&& \frac{1}{2\kappa^2} \int d^{10}x (-G)^{1/2} R - 
\frac{1}{4\kappa^2} \int \biggl( d\phi \wedge *d\phi + g_s 
e^{-\Phi}H_3 \wedge \star H_3 
\nonumber \\
&&+ g_s^{1/2} e^{3\phi/2}F_2 \wedge \star F_2  + g_s^{3/2} e^{\phi/2} \tilde F_4 \wedge \star \tilde F_4 + g_s^2 B_2 \wedge F_4 \wedge F_4 \biggr),
\end{eqnarray}
where 
\begin{equation}
\tilde F_4 = F_4 - C_1 \wedge H_3 \;\;\; , \;\;\; F_4 = dC_3 \;\;\; ,
\;\;\; F_2 = dC_1 \ , 
\end{equation}
and $2\kappa^2 = (2\pi)^7 {\alpha'}^4 g_s^2$.

We then make an ansatz for the metric and field strengths that 
depends on only one coordinate, namely the radius.
The metric is
\be
ds_{10}^2 = e^{-5z} dx^\mu dx^\nu \eta_{\mu\nu}
+ e^{3z} ds_7^2
\ee
where $ds_7^2$ is as in (\ref{g2metric}).
Thus the metric will depend on three
undetermined functions $z(r)$, $h(r)$,
$a(r)$, and $b(r)$.  For convenience,
we redefine
\be
e^{u(r)} = a(r) \; ; \; \; \; e^{v(r)} = b(r) \ .
\ee

The field strengths are then chosen to
obey the Bianchi identities.  The
two form $F_2=0$.  
The dilaton is left as $\phi(r)$.  
The other forms are
\be
\ell B_2 = m \left( g_1(r) X_2 + g_2(r) J_2 \right) \ ,
\ee
and
\begin{eqnarray}
g_s F_4 &=& K(r) d^3x \wedge dr + 2m(g_3(r) + c_1) J_2 \wedge J_2
+ 2m(g_3(r) + c_2) X_2 \wedge J_2 \nonumber \\
&& + m g_3(r)' \epsilon_{ijk} \mu^i dr \wedge D\mu^j \wedge J^k \ 
\end{eqnarray}
where $c_1$ and $c_2$ are integration constants.

After some algebra, the following one dimensional effective
action emerges
\be
S = \frac{\ell^5 \Vol(M^{2,1}) \Vol(M_6)}{2\kappa^2} \int {\mathcal L} \, dr
\ee
where $M^{2,1}$ is the 2+1 dimensional Minkowski space and the 
$M^6$ are the level surfaces of the seven dimensional
$G_2$ manifold.  The Lagrangian is ${\mathcal L} = T-V$
where
\begin{eqnarray}
T &=& \frac{e^{2u+4v}}{h} \left[
-30(z')^2 + 2(u')^2 + 16u'v' + 12(v')^2 - \frac{1}{2} (\phi')^2
\right. \nonumber \\
&& -\frac{1}{2} \frac{m^2}{\ell^6} g_s e^{-\phi-6z} 
\left( (g_1')^2 e^{-4u} + 2 (g_2') e^{-4v} \right) 
\nonumber \\
&& \left.
-2 \frac{m^2}{\ell^6} g_s^{-1/2} e^{\phi/2 - 9z -2u-4v} 
(g_3')^2 \right]
\end{eqnarray}
and
\begin{eqnarray}
V &=& h e^{-2u-4v} \left[
2e^{2u+8v} - 2e^{6u+4v} + 12 e^{4u+6v} -
2 \frac{m^2}{\ell^6} g_s e^{-\phi-6z+2u+4v} (g_1 + g_2)^2 \right.
\nonumber \\
&&\left.
- 4 \frac{m^2}{\ell^6} g_s^{-1/2} e^{\phi/2-9z}
\left( 2(g_3 + c_1)^2 e^{4u} + (g_3 + c_2)^2 e^{4v} \right )
\right] \nonumber \\
&&- \frac{1}{2} g_s^{-1/2} e^{\phi/2+15z+2u+4v} h^{-1} K^2 +
\frac{4m^2}{\ell^6} K \left[ g_3 (g_1 + g_2) + g_1 c_1 + g_2 c_2 \right] \ .
\label{poto}
\end{eqnarray}
Notice that the functions $h(r)$ and $K(r)$ are non-dynamical.
We leave $h(r)$ in the Lagrangian in order to retain
reparametrization invariance.  
After all, we
eventually 
hope to identify $r$ with the radius of section 2.
The Lagrangian simplifies a little if we solve
for the minimum value of $K(r)$ and replace
the old Lagrangian with a new one evaluated
at this minimum:
\be
K(r) = \frac{4m^2}{\ell^6} g_s^{1/2} e^{-\phi/2-15z -2u-4v} h 
\left[ g_3 (g_1 + g_2) + g_1 c_1 + g_2 c_2 \right] \ .
\ee
Having made this replacement, the potential becomes
\begin{eqnarray}
V \rightarrow - h e^{-2u-4v} \left[
2 e^{2u+8v} - 2 e^{6u+4v} + 12 e^{4u+6v} -
2 \frac{m^2}{\ell^6} g_s e^{-\phi-6z+2u+4v} (g_1 + g_2)^2 \right.
\nonumber \\
\left.
- 4 \frac{m^2}{\ell^6} g_s^{-1/2} e^{\phi/2-9z}
\left( 2(g_3 + c_1)^2 e^{4u} + (g_3 + c_2)^2 e^{4v} \right )
\right] \nonumber \\
+ \frac{8m^2}{\ell^{12}} g_s^{1/2} e^{-\phi/2 -15z-2u-4v}
h \left[
g_3 (g_1 + g_2) + g_1 c_1 + g_2 c_2 
\right]^2
\ .
\label{pot}
\end{eqnarray}

This Lagrangian can be written in terms of a superpotential.
Indeed, let
\be
T = G_{ab} \dot \varphi^a \dot \varphi^b
\ee
where the $\varphi^a$ correspond to the functions
$u$, $v$, $z$, $\phi$, and $g_i$, and 
the dot means differentiation with respect
to the radius $r$.  
The $h'$, as $h$ is nondynamical, does not
appear in $T$. 
In this notation, 
the potential can be written as
\be
V = -G^{ab} \frac{\partial W}{\partial \varphi^a} 
\frac{\partial W}{\partial \varphi^b}
\ee
where $G^{ab}$ is the inverse of $G_{ab}$ and 
\be
W = 2 \left[ e^{u+4v} + e^{3u+2v} - \frac{m^2}{\ell^6}
g_s^{1/4} e^{-\phi/4 -15z/2} 
\left( g_3 (g_1 + g_2) + c_1 g_1 + c_2 g_2 \right)
\right] \ .
\ee

The existence of the superpotential $W$ implies that
\be
{\mathcal L} = G_{ab} 
\left( \dot \varphi^a - G^{ac} \frac{\partial W}{\partial \varphi^c} \right)
\left( \dot \varphi^b - G^{bd} \frac{\partial W}{\partial \varphi^d} \right)
+ 2 \frac{dW}{dr} \ .
\ee
In this way, a simple set of first order, differential equations emerges,
\be
\dot \varphi^a = G^{ab} \frac{\partial W}{\partial \varphi^b} \ .
\ee
It is straightforward now to make sure that the metric
functions and field strengths of  
section 2 satisfy this system of first order
equations.  One easy observation is that
$2z' = \phi'$.  This
observation is equivalent to the statement that
the warp factor 
$H(r)^{1/4} \equiv e^{2z} = g_s^{-1} e^{\phi(r)}$
noted earlier.  With this
substitution, the first order system becomes:
\begin{eqnarray}
u'/h &=&  \left( e^{-u} - e^{u-2v} \right) \ , \\
v'/h &=& e^{2u-2v} \ , \\ 
z'/h &=& -\frac{m^2}{2 \ell^6} e^{-2u-4v-8z} 
\left( g_3(g_1 + g_2) + c_1 g_1 + c_2 g_2 \right) 
\label{warpfirst}
\ , \\
g_1'/h &=& 4 (g_3 + c_1) e^{2u-4v} \label{g1} \ ,\\
g_2'/h &=& 2 (g_3 + c_2) e^{-2u} \label{g2} \ , \\
g_3'/h &=& (g_1 + g_2) \ .
\end{eqnarray} 
As promised, the $h$ allows us some freedom in the choice of
$r$.  It is convenient to choose $h(r)$ as in (\ref{hab}).
It is then straightforward to see that $a(r)$ and $b(r)$ of
section 2 satisfy the first pair of equations 
in the first order system.

The last four differential equations have a simple interpretation.
The last three differential equations constitute a geometric
condition on the forms.  They enforce the condition
that $G_3 = \star_7 G_4$ where $\star_7$ is the
Hodge dual with respect to the $G_2$ manifold.
As discussed in the text, 
the equation for the warp factor (\ref{warpfirst}) 
corresponds to
the integrated equation of motion for the four form field
strength $F_4$.

We present a few details necessary to
integrate (\ref{warpfirst}). 
The relations between the $f_i$ of the previous
section and the $g_i$ are as follows:
\be
g_1' = f_1 \; ; \; \; \; 
g_2' = f_2 \; ; \; \; \;
g_3' = h f_3 \; .
\ee
Now the $f_i$ depend on elliptic integrals and
the reader might expect that the $g_i$ can at best
be expressed as double integrals.  It is
surprising but true that the $g_i$ are no
more complicated than the $f_i$.  Indeed
\be
g_1 = -8 a^2 b^2 f_3 \; ; \; \; \;
g_2 = (1+8a^2b^2) f_3 \ .
\ee
The function $g_3$ can be expressed in terms
of $f_1$ or $f_2$ using either 
(\ref{g1}) or (\ref{g2}), giving a relation
between the integration constants
$32(c_1 - c_2) = 3$.  (Note that
we are free to change $g_3$ by a constant
without changing the solution; the 
geometry does not
specify both $c_1$ and $c_2$, only
the difference.)


After some algebra, we find that
\be
H(r) = \frac{m^2}{2 \ell^6} \int_r^\infty \rho 
\left( 2u_2(\rho) u_3(\rho) - 3u_3(\rho) \right) d\rho \ .
\ee

\end{appendix}

\end{document}